\documentclass[review]{elsarticle}
\usepackage[utf8]{inputenc}
\usepackage{amssymb,amsmath,array}
\usepackage{amsthm}
\usepackage{fixltx2e}
\usepackage{pdfpages}
\usepackage{float}
\usepackage[caption = false]{subfig}

\usepackage{todonotes}

\newcommand{\argmin}{\operatornamewithlimits{argmin}}

\begin{document}

\begin{frontmatter}

\title{Exploiting sparsity to build efficient kernel based collaborative filtering for top-N item recommendation}

\author{Mirko Polato}
\address{University of Padova - Department of Mathematics\\ Via Trieste, 63, 35121 Padova - Italy}

\author{Fabio Aiolli}
\address{University of Padova - Department of Mathematics\\ Via Trieste, 63, 35121 Padova - Italy}

\begin{abstract}
The increasing availability of implicit feedback datasets has raised the interest in developing effective collaborative filtering techniques able to deal asymmetrically with unambiguous positive feedback and ambiguous negative feedback. 

In this paper, we propose a principled kernel-based collaborative filtering method for top-N item recommendation with implicit feedback. 
We present an efficient implementation using the linear kernel, and we show how to generalize it to  kernels of the dot product family preserving the efficiency.

We also investigate on the elements which influence the sparsity of a standard cosine kernel. This analysis shows that the sparsity of the kernel strongly depends on the properties of the dataset, in particular on the long tail distribution.
We compare our method with state-of-the-art algorithms achieving good results both in terms of efficiency and effectiveness.
\end{abstract}


\end{frontmatter}

\section{Introduction}\label{intro}

Collaborative filtering (CF) techniques can make recommendation to a user exploiting information provided by similar users.
The typical CF setting consists of a set $\mathcal{U}$ of $n$ users, a set $\mathcal{I}$ of $m$ items, and the so-called rating matrix $\mathbf{R} = \{r_{ui}\} \in \mathbb{R}^{n \times m}$. Ratings represent, numerically, the interactions between users and items. These interactions can be of two different types: explicit feedbacks and implicit feedbacks. The former are interactions performed by users in a direct way, such as, by giving a one-to-five star rate, thumbs-up versus thumbs-down, and so on. The latter, instead, are actions performed by users without the awareness of giving some kind of feedback to the system, e.g., clicks, views, elapsed time on a web page and so on.

The explicit setting have got most of the attention from the CF research community and a lot of rating prediction algorithms have been proposed. More recently however, the focus is constantly shifting towards the implicit feedback scenario since users are not always willing to give their opinion explicitly.
Furthermore, implicit interactions are easier to collect: for an already existing system, it is sufficient to store the operations done by a user in a session without changing the front-end and consequently there are no additional burden for the user.

One problem arising in implicit feedback is that typically the feedback is asymmetric, that is, while there can be evidence that a certain user interacted and hence showed some interest towards a product (unambiguous feedback), the opposite is not true, that is the missing evidence of interaction does not imply that a given user dislikes that item.

In this paper we focus on the implicit feedback scenario, and so we assume binary ratings, $r_{ui} \in \{0,1\}$, where $r_{ui} = 1$ means that user $u$ interacted with the item $i$ (unambiguous feedback) and $r_{ui} = 0$ means there is no evidence that user $u$ interacted with the item $i$ (ambiguous feedback).

Specifically, the main contribution of this paper is a CF framework for top-N item recommendation based on the seminal work described in \cite{Aiolli:2014gx}.
Starting from the original convex optimization problem, we propose an efficient variation of that formulation which makes the algorithm able to manage very large scale datasets, by preserving the effectiveness of the original algorithm. This new formulation is then extended to be used with general kernels thus augmenting the expressiveness of the representation in collaborative filtering.

It is well known that kernels are not suited for large datasets since they have a prohibitive computational complexity. Datasets used in CF domains are usually large: rating matrices $\mathbf{R}$ have tens of thousands of rows and columns and so the application of kernels seems to be difficult.
Nevertheless, we can observe that the complexity of the computation of a kernel strictly depends on its sparsity and hence being able to control the sparsity makes the application of kernel methods to very large datasets feasible.
Unfortunately, kernels are usually dense unless we are working on very large and sparse feature spaces and this is in contrast with what we would like to have.

Exploiting a well known result from harmonic theory \cite{AISTATS2012_KarK12}, we are able to demonstrate how using \emph{dot product kernels}, a generalization of normalized homogeneous polynomial kernels, kernels as sparse as the standard cosine kernel can be obtained without changing the solution of the problem.

However, it should be noted that there is no guarantees about the sparsity of a cosine kernel because it is closely related to the distribution of the non-zero values in the matrix $\mathbf{R}$.
CF datasets are known to be very sparse, and also they are usually extracted from e-services which makes them subject to the long tail phenomenon that is often associated with the power law distribution.

Even if not every long tail is a power law \cite{site:singers}, this tailed distribution over the item ratings in $\mathbf{R}$ poses a strong bias on the density of the resulting kernel. In fact, assuming the long tail, in $\mathbf{R}$ there will be few dense columns and many sparse ones, which intuitively leads to a rather sparse kernel. On the other side, a long tailed distribution over the user ratings can lead to dense kernels matrices since different items tend to be rated by mostly the same users.
We provide a deep analysis of this phenomenon showing theoretically and empirically which are the conditions for which a dataset is likely to produce a sparse dot product kernel.
It is worth to notice that these results are not only applicable to CF contexts, but they apply on every other context where the data distribution is long tailed.\\

Finally, in the experimental section we compare our framework with two state-of-the-art methods in top-N recommendation. The empirical work shows how our method can achieve good results in terms of AUC (Area Under the ROC Curve) and also in terms of efficiency.\\

Summarizing, the contribution of this paper is $3$-fold:
\begin{enumerate}
\item Starting from the seminal work described in \cite{Aiolli:2014gx} we propose an optimized CF framework for top-N recommendation. Specifically, our proposed method results far more efficient of the original enabling the application of the method to large and very large datasets (e.g. in the order of 1 million of users/items and 50 million of rates) while preserving the state-of-the-art effectiveness of the original algorithm.
\item The formulation depicted above is  generalized to be used with   non-linear kernels as dot-product kernels, that is kernel functions in the form $k(\mathbf{x},\mathbf{y}) = f(\mathbf{x} \cdot \mathbf{y})$ where $f$ is a non-linear function admitting a Maclaurin expansion with non-negative coefficients. A sparsification method for dot-product kernels is also provided making the sparsity of any dot-product kernel equal to the sparsity of the linear or cosine kernel.
\item A theoretical and empirical analysis concerning the sparsity of the standard cosine kernel is proposed. This analysis shows that the sparsity of the kernel produced depends on the properties of the user activity and item popularity long-tails. In particular, long-tails observed on item popularity improve the sparsity of the kernels while long-tails observed on user activity is detrimental for the sparsity of the kernels.
\end{enumerate}

This paper is an extended version of a preliminary paper presented at ESANN 2016 \cite{polato2016esann}. In particular, preliminary work about the contribution 1 and 2 above was already presented in that work while contribution 3 is novel. Moreover, the extensive experimental work of this paper was not present in the preliminary version.

The rest of the paper is organized as follows. In Section \ref{notation} we will introduce the notation used throughout the paper. Section \ref{related} presents related works on top-N recommendation for implicit feedback. Sections \ref{cfomd} and \ref{cfkomd} describes our framework with a particular focus on the applicability of our proposed kernel method.
Finally, Sections \ref{result} and \ref{conclusion} show the experimental results and which directions our research can follow in the future.\\

\section{Notation}\label{notation}
In this section we provide some useful notation used throughout the paper. Recommender algorithms are thought to give suggestions to users about items. 
We call the set of users as $\mathcal{U}$ such that $|\mathcal{U}|=n$, the set of items as $\mathcal{I}$ such that $|\mathcal{I}|=m$ and the set of ratings $\mathcal{R} = \{(u,i)\}$.

We refer to the binary rating matrix with $\mathbf{R} = \{r_{ui}\} \in \mathbb{R}^{n \times m}$, where users are on the rows and items on the columns. 
We add a subscription to both user and item sets to indicate, respectively, the set of items rated by a user $u$ ($\mathcal{I}_u$) and the set of users who rated the item $i$ ($\mathcal{U}_i$). 

\section{Related works}\label{related}

Top-N recommendation finds application in many different domains such as TV and movies \cite{koren:2008}, books \cite{Aiolli:2014gx}, music \cite{Aiolli:2013dp,Tan:2011}, social media \cite{Wang:2013} and so on. 

Top-N recommendation methods can be divided into two macro categories. 

The first is the neighbourhood-based CF algorithms \cite{karypis:2001}, in which the recommendation for a target user is made by using the ratings of the most similar users. This category comprises the so called memory-based methods that do not need the construction of a model, but they directly use the data inside the rating matrix.

Despite, in general, these methods suffer from low accuracy, in 2013 the winner of the remarkable challenge organized by Kaggle, the Million Songs Dataset challenge \cite{McFee:2012:MSD:2187980.2188222}, was an extension of the well known item-based nearest-neighbors (NN) algorithm \cite{karypis:2004}.
This extension \cite{Aiolli:2013dp} (here called MSDW) introduced an asymmetric similarity measure, called asymmetric cosine. 
In a classic item-based CF method, the scoring function for a user-item pair $(u,i)$ is computed by a weighted sum over the items liked by $u$ in the past, that is:
$$
	\hat{r}_{ui} = \sum\limits_{j \in \mathcal{I}} w_{ij} r_{uj} = \sum\limits_{j \in \mathcal{I}_u} w_{ij},
$$
where $w_{ij}$ expresses the similarity between item $i$ and item $j$. 

One of the main contribution, presented in \cite{Aiolli:2013dp}, is the asymmetric cosine (asymC) similarity. The intuition behind asymC comes from the observation that the cosine similarity can be expressed, over a pair of items $(a,b)$, as the square root of the product of the reciprocal conditional probabilities. Let $\mathbf{a}, \mathbf{b} \in \{0,1\}^n$ be respectively the binary vector representations of items $a$ and $b$.
The idea of asymmetric cosine similarity is to give different weights to the conditional probabilities, that is
$$
	S_{\alpha}(a, b) = \frac{\mathbf{a}^\top \mathbf{b}}{\|\mathbf{a}\|^{2\alpha} \|\mathbf{b}\|^{2(1-\alpha)}} = P(a|b)^{\alpha} P(b|a)^{1-\alpha},
$$
with $0 \leq \alpha \leq 1$. In case of binary rates this asymmetric similarity can be computed as in the following. Let $\mathcal{U}_i$ represents the set of users who rated the item $i$, then the asymC between item $i$ and item $j$ is defined by:
$$
	w_{ij} = S_{\alpha}(i,j) = \frac{|\mathcal{U}_i \cap \mathcal{U}_j|}{|\mathcal{U}_i|^{\alpha}|\mathcal{U}_j|^{1-\alpha}}.
$$
Besides its outstanding performance in terms of mAP@500, the MSD winning solution is also easily scalable to very large datasets. However, one drawback of this solution is that it is not theoretically well founded. 

The second category is the model-based CF techniques, which construct a model of the information contained in the rating matrix $\mathbf{R}$. In the last two decades many different model-based approaches have been proposed. A particular attention has been devoted to latent factor models which try to factorize the rating matrix into two low-rank matrices, $\mathbf{R}=\mathbf{W}\mathbf{X}$, which represent user-factors ($\mathbf{W}$) and item-factors ($\mathbf{X}$). These factors are ``meta-features'' that define the user tastes and how much of these features represent an item. Usually these methods are referred to as matrix factorization methods.
The prediction for a user-item pair is simply done by a dot product of the corresponding row and column in the factorized matrices. 

One of the most used matrix factorization approach for implicit feedback is presented in \cite{koren:2008} (WRMF: Weighted Regularized Matrix Factorization). In this work Hu et al. propose an adaptation of the classic SVD (Singular Value Decomposition) method in which they minimize the square-loss using two regularization terms in order to avoid overfitting. Their optimization criterion is defined as:
$$
	\sum\limits_{u\in\mathcal{U}}\sum\limits_{i\in\mathcal{I}} c_{ui} (\mathbf{w}_u^\top\mathbf{x}_i - 1)^2 + \lambda\|\mathbf{W}\|^2 + \lambda\|\mathbf{X}\|^2,
$$
where $c_{ui}$ are a-priori weights for each pair $(u,i)$ such that positive feedbacks have higher weights. Actually, this method uses information about the rating values (not binary) to give more importance to user-item interactions with high rating values, in fact, $c_{ui}$ is calculated by: $c_{ui} = 1 + \alpha r_{ui}$. In their experiments the best performances has been achieved with $\alpha = 40$, but with binary rating matrices this parameter losses a lot of its importance. 
As we will see in the experimental section, changing the value of $\alpha$ does not change the performance of the algorithm.
A similar approach has been used by Wang et al. \cite{Wang:2016} where they proposed a framework for broadcast email prioritization based on a novel active learning method.

In \cite{rendle:2009} the top-N recommendation problem has been formulated as a ranking problem. Rendle et al. proposed a Bayesian Personalized Ranking (BPR) criterion that is the maximum posterior estimator derived from a Bayesian analysis. They also provide a learning method based on stochastic gradient descent with bootstrap sampling. They finally show how to adopt this criterion for kNN (BPRkNN) and MF methods (BPRMF).

In \cite{karypis:2011} Ning et al. presented SLIM (Sparse LInear Method) which generates top-N recommendation by aggregating from user rating profiles. SLIM learns an aggregation coefficient matrix by solving a regularized optimization problem.

More recently, a new principled algorithm for CF, which explicitly optimizes the AUC, has been proposed with very nice performances on the MovieLens dataset \cite{Aiolli:2014gx}. Since this algorithm represents the seminal work for our framework, we will discuss about it in details in the next section.

A more direct approach in order to build a good ranking over the items is the so called \textit{learning to rank}\cite{Cao:2007}. Learning to rank methods exploit supervised machine learning to solve ranking problems. These techniques can be divided into three categories: pointwise approaches \cite{Wang:2013,Burges:2005,Ostuni:2013} in which for each query-document (i.e., user-item) pair a score is predicted and then used to build the ranking; pairwise approaches \cite{Zhong:2014,Rendle:2014} face the ranking problem as a binary classification one (positive document versus negative ones) in which they try to minimize the number of inversions in the ranking; listwise approaches \cite{Shi:2010,Huang:2015} try to directly optimize one of the ranking evaluation measures. The big challenge here is the fact that most of the measures are not continuous function w.r.t. the model's parameters and for this reason approximations have to be used.

\section{CF-OMD framework}\label{cfomd}
\subsection{CF-OMD}
In this section we present the seminal CF algorithm, called CF-OMD (Optimization of the Margin Distribution) \cite{Aiolli:2014gx}, for top-N recommendation inspired by preference learning \cite{aiolli:2005}\cite{Aiolli:2008}, and designed to explicitly maximize the AUC (Area Under the ROC Curve).
 
Consider the matrix $\mathbf{W} \in \mathbb{R}^{n \times k}$ be the embeddings of users in a latent factor space and  $\mathbf{X} \in \mathbb{R}^{k \times m}$ be the embeddings of items in the space.
Given a user, a ranking over items can be induced by the factorization $\hat{\mathbf{R}} = \mathbf{W}\mathbf{X}$, where $\hat{r}_{ui} = \mathbf{w}_u^\top \mathbf{x}_i$ with the constraint $\|\mathbf{w}_u\| = \|\mathbf{x}_i\| = 1$. The model parameters (i.e., $\mathbf{W}, \mathbf{X}$) are generally computed minimizing a regularized loss function over the ground truth matrix $\mathbf{R}$.
Let now fix the item representation as $\mathbf{x}_i = \mathbf{r}_i/\|\mathbf{r}_i\|$, 
and let $\rho(i \prec_u j)=(\hat{r}_{ui}-\hat{r}_{uj})/2 = \mathbf{w}_u^\top(\mathbf{x}_i - \mathbf{x}_j)/2$ be the margin for an item pair $(i,j)$ for user $u$. 
Let also define the probability distribution over the positive and negative items for $u$, 
$$
\mathbf{A}_u = \{\boldsymbol\alpha_u \in \mathbb{R}_+^m | \sum_{i \in \mathcal{I}_u} \alpha_{ui} = 1, \sum_{i \notin \mathcal{I}_u} \alpha_{ui} = 1\}.
$$
In \cite{Aiolli:2014gx} it is proposed an approach to maximize the minimum margin inspired by preference learning where the ranking task is posed as a two-player zero-sum game. Let $P_{\textit{max}}$ and $P_{\textit{min}}$ be the players: on each round of the game, $P_{\textit{min}}$ picks a preference $i \prec_u j$ and, simultaneously, $P_{\textit{max}}$ picks an hypothesis $\mathbf{w}_u$ with the aim of maximizing the margin $\rho(i \prec_u j)$.
The value of the game, i.e., the expected margin, is computed by:  
$$
	\mathbb{E}_\alpha[\rho] = \frac{1}{2} \mathbf{w}_u^\top \mathbf{X} \mathbf{Y}_u \boldsymbol\alpha_u,
$$
where $\mathbf{Y}_u$ is a diagonal matrix, $\mathbf{Y}_u = \textit{diag}(\mathbf{y}_u)$, such that $y_{ui} = 1$ if $i \in \mathcal{I}_u$, $-1$ otherwise.
It can be demonstrated that the $\mathbf{w}^*_u$ maximizing the expected margin is equal to $\mathbf{w}_u^* = \mathbf{X}\mathbf{Y}_u\mathbf{\boldsymbol{\alpha}}_u$ normalized.
Finally, the best strategy for $P_{\textit{min}}$ can be expressed as a convex quadratic optimization problem:
\begin{equation}\label{opt1}
	\boldsymbol{\alpha}_u^* = \argmin\limits_{\boldsymbol{\alpha}_u \in \mathbf{A}_u} \;\;\boldsymbol{\alpha}_u^\top \left( \mathbf{Y}_u \mathbf{X}^\top \mathbf{X} \mathbf{Y}_u + \boldsymbol\Lambda \right) \boldsymbol{\alpha}_u,
\end{equation}
in which $\boldsymbol{\Lambda}$ is a diagonal matrix such that $\boldsymbol{\Lambda}_{ii} = \lambda_p$ if $i \in \mathcal{I}_u$, otherwise $\boldsymbol{\Lambda}_{ii} = \lambda_n$, where $\lambda_p$ and $\lambda_n$ are regularization parameters $(\lambda_p, \lambda_n \geq 0)$.

Although this algorithm has shown state-of-the-art results in terms of AUC, it is not suitable to deal with large datasets. In fact, let assume that each optimization problem can be solved by an algorithm with a complexity quadratic on the number of parameters. 
Then the global complexity would be $O(n_{\textit{ts}} m^2)$, where $n_{\textit{ts}}$ is the number of users in the test set, and for the MSD it would be $O(10^{19})$.

\subsection{Efficient CF-OMD}\label{eff_cf-omd}
The main issue of CF-OMD, in terms of efficiency, is the number of parameters which is equal to the cardinality of the item set.
Analyzing the results reported in \cite{Aiolli:2014gx}, we noticed that high values of $\lambda_n$ did not particularly affect the results, because it tends to flatten the contribution of the ambiguous negative feedbacks toward the average, mitigating the relevance of noisy information.

In CF contexts the data sparsity is particularly high, this means, on average, that the number of ambiguous negative feedbacks is orders of magnitude greater than the number of positive feedbacks. Formally, given a user $u$, let $m_u^+ = |\mathcal{I}_u|$ and $m_u^- = |\mathcal{I} \setminus \mathcal{I}_u|$ then $m = m_u^- + m_u^+$, where $m_u^+ \ll m_u^-$, and generally $O(m) = O(m_u^-)$.

On the basis of this observation, we can simplify the optimization problem (\ref{opt1}), by fixing $\lambda_n = +\infty$, which means that $\forall i \notin \mathcal{I}_u, \alpha_{ui} = 1/m_u^-$:
\begin{align}
		\boldsymbol{\alpha}_u^* &= \argmin\limits_{\boldsymbol{\alpha}_u} \: \|\boldsymbol\alpha_{u^+}^\top \mathbf{X}_{u^+} - \boldsymbol\mu_u^-\|^2 + \lambda_p \|\boldsymbol\alpha_{u^+}\|^2 \\
		&= \argmin\limits_{\boldsymbol{\alpha}_u} \: \|\boldsymbol\alpha_{u^+}^\top \mathbf{X}_{u^+} \|^2 - \|\boldsymbol\mu_u^-\|^2 -2\boldsymbol{\alpha}^\top_{u^+}\mathbf{X}_{u^+}^\top\boldsymbol{\mu}_{u}^- + \lambda_p \|\boldsymbol\alpha_{u^+}\|^2 \\
		&= \argmin\limits_{\boldsymbol{\alpha}_{u^+} \in \mathbf{A}_u} \;\;\boldsymbol{\alpha}_{u^+}^\top \mathbf{X}_{u^+}^\top \mathbf{X}_{u^+} \boldsymbol{\alpha}_{u^+} + \lambda_p\|\boldsymbol{\alpha}_{u^+}\|^2 - 2\boldsymbol{\alpha}^\top_{u^+}\mathbf{X}_{u^+}^\top\boldsymbol{\mu}_{u}^-, \label{opt2}
\end{align}
where
$$
\boldsymbol{\mu}_{u}^- = \frac{1}{m_u^-} \sum_{i \notin \mathcal{I}_u} \mathbf{x}_i
$$ 
is the centroid of the convex hull spanned by the negative items and $\boldsymbol{\alpha}_{u^+}$ are the probabilities associated with the positive items, $\mathbf{X}_{u^+}$ is the sub-matrix of $\mathbf{X}$ containing only the columns corresponding to the positive items.
The number of parameters in (\ref{opt2}) is $m_u^+$ and hence the complexity from $O(n_{\textit{ts}} m^2)$ drops to $O(n_\textit{ts} {\overline{m}_u^+}^2)$, where $\overline{m}_u^+ = \mathbb{E}[|\mathcal{I}_u|]$ is the expected cardinality of the positive item set. In MSD $\overline{m}_u^+ \approx 47.46$ which leads to a complexity $O(10^8)$.

\subsubsection{Implementation trick}\label{impl}
Notwithstanding the huge improvement in terms of complexity, a na\"{i}ve implementation would have an additional cost due to the calculation of $\boldsymbol{\mu}_{u}^-$. For all users in the test set the cost would be $O(n_{\textit{ts}} n \overline{m}_u^-)$, where $\overline{m}_u^- = \mathbb{E}[|\mathcal{I} \setminus \mathcal{I}_u|]$, and it can be approximated with $O(n_{\textit{ts}}nm)$.

To overcome this bottleneck, we propose an efficient incremental way of calculating $\boldsymbol{\mu}_{u}^-$. Consider the mean over all items 
$$
\boldsymbol{\mu} = \frac{1}{m} \sum_{i \in \mathcal{I}} \mathbf{x}_i,
$$ 
then, for a given user $u$, we can express 
$$
\boldsymbol{\mu}_{u}^- = \frac{1}{m_u^-} \left( m \cdot \boldsymbol{\mu} - \sum_{i \in \mathcal{I}_u} \mathbf{x}_i \right).
$$
From a computational point of view, it is sufficient to compute the sum $\sum_{i \in \mathcal{I}} \mathbf{x}_i$
 once (i.e., $m \cdot \boldsymbol{\mu}$) and then, for every $\boldsymbol{\mu}_u^-$, subtract the sum of the positive items. Using this simple trick, the overall complexity drops to $O(nm) + O(n_{\textit{ts}}^2\overline{m}_u^+)$.

In the experimental section we successfully applied this algorithm to the MSD achieving competitive results against the state-of-the-art method but with higher efficiency.

\section{Kernelized CF-OMD}\label{cfkomd}
The method proposed in Section \ref{eff_cf-omd}, can be seen as a particular case of a kernel method. In fact, $\mathbf{X}_{u^+}^\top \mathbf{X}_{u^+}$ is a kernel matrix, let call it $\mathbf{K}_{u^+}$ with the corresponding (linear) kernel function $K: \mathbb{R}^{m_u^+} \times \mathbb{R}^{m_u^+} \rightarrow \mathbb{R}$. Given $K$ we can reformulate (\ref{opt2}) as:
\begin{equation}\label{opt3}
	\boldsymbol{\alpha}_{u^+}^* = \argmin\limits_{\boldsymbol{\alpha}_{u^+} \in \mathbf{A}_u} \;\;\boldsymbol{\alpha}_{u^+}^\top \mathbf{K}_{u^+} \boldsymbol{\alpha}_{u^+} + \lambda_p\|\boldsymbol{\alpha}_{u^+}\|^2 - 2\boldsymbol{\alpha}_{u^+}^\top\mathbf{q}_u,
\end{equation}
where elements of the vector $\mathbf{q}_u \in \mathbb{R}^{{m_u^+}}$ are defined as 
$$
\quad q_{ui} = \frac{1}{m_u^-} \sum_{j \notin \mathcal{I}_u} K(\mathbf{x}_i, \mathbf{x}_j).
$$

Actually, inside the optimization problem (\ref{opt3}) we can plug any kernel function. Throughout the paper we will refer to this method as CF-KOMD. Generally speaking, the application of kernel methods on a huge dataset have an intractable computational complexity.
Without any shrewdness the proposed method would not be applicable because of the computational cost of the kernel matrix and $\mathbf{q}_u$.

An important observation is that the complexity is strictly connected with the sparsity of the kernel matrix which is, unfortunately, commonly dense. 
However, we can leverage on a well known result from harmonic theory \cite{AISTATS2012_KarK12} to keep the kernel as sparse as possible without changing the solution of CF-KOMD. 
\newtheorem{theorem}{Theorem}
\begin{theorem}
A function $f:\mathbb{R}\rightarrow \mathbb{R}$ defines a positive definite kernel $k : \mathbf{B}(0, 1) \times \mathbf{B}(0,1)$ as $k: (\mathbf{x}, \mathbf{y}) \mapsto f(\mathbf{x}\cdot \mathbf{y})$ iff $f$ is an analytic function admitting a Maclaurin expansion with non-negative coefficients, $f(x) = \sum_{s=0}^\infty a_s x^s, a_s \geq 0$.
\end{theorem}

As emphasized in \cite{AISTATS2012_KarK12, aiolli:2017}, many kernels used in practice \cite{Scholkopf:2001} satisfy the above-mentioned condition. These kernels are called dot product kernels because they are defined as a function of the dot product of the input vectors. Table \ref{tab:dpk} gives some example of these kind of kernels.

\begin{table}[h!]
	\centering
	\renewcommand{\arraystretch}{2}%
	\begin{tabular}{|c|c|c|}
		\hline
		\textbf{Kernel} & \textbf{Definition} & \textbf{Coefficients} $a_s$ \\ \hline
		Linear & $\mathbf{x}\cdot\mathbf{y}$ & $1, s=1$ \\ \hline
		Polynomial & $(\mathbf{x}\cdot\mathbf{y} + c)^d$ & ${d \choose n} c^{(d-s)}, \forall s \in [0,d]$ \\ \hline
		RBF & $e^{-\gamma\|\mathbf{x}-\mathbf{y}\|^2}$ & $e^{-2\gamma \frac{(2\gamma)^{2s}}{s!}}, \forall s$ \\ \hline
		Tanimoto & $\frac{\mathbf{x}\cdot\mathbf{y}}{\|\mathbf{x}\|+\|\mathbf{y}\|-\mathbf{x}\cdot\mathbf{y}}$& $2^{-s}, \forall s > 0$ \\ \hline
	\end{tabular}
	\caption{Some examples of dot product kernels.\label{tab:dpk}}
\end{table}

We can observe that the kernel matrices induced by these kernels are, in general, dense due to the zero degree term (i.e., $s = 0$) which is a constant added to all the entries.
Adding a constant to a whole matrix means a space translation and we can demonstrate that this operation does not affect the margin in CF-KOMD (this is valid also in the generic formulation of CF-OMD).
\begin{proof}
Let $\mathbf{K} = \mathbf{K}_0 + \hat{\mathbf{K}}$ be a dot product kernel matrix where $\mathbf{K}_0$ is the constant matrix induced by the 0 degree term of the MacLaurin expansion (i.e. $s = 0$). Let also $\mathbf{q}_u$ be consequently defined as:
\begin{align}
	\quad q_{ui} &= \frac{1}{m_u^-} \sum_{j \notin \mathcal{I}_u} \left( K_0(\mathbf{x}_i, \mathbf{x}_j) + \hat{K}(\mathbf{x}_i, \mathbf{x}_j) \right) \\
	&= \frac{1}{m_u^-} \left[ m_u^- \cdot k_0 + \sum_{j \notin \mathcal{I}_u} \hat{K}(\mathbf{x}_i, \mathbf{x}_j) \right ] \\
	&= k_0 + \hat{q}_{ui} \Rightarrow \mathbf{q}_u = \mathbf{k}_0 + \hat{\mathbf{q}}_u.
\end{align}

We can rewrite the optimization problem	(\ref{opt3}) as (we omit the $u^+$ subscription for brevity):
\begin{align}
	\boldsymbol{\alpha}^* &= \argmin\limits_{\boldsymbol{\alpha} \in \mathbf{A}_u} \;\;\boldsymbol{\alpha}^\top (\mathbf{K}_0 + \hat{\mathbf{K}}) \boldsymbol{\alpha} + \lambda_p\|\boldsymbol{\alpha} \|^2 - 2\boldsymbol{\alpha}^\top(\mathbf{k}_0 + \hat{\mathbf{q}}_u) \\
	&= \argmin\limits_{\boldsymbol{\alpha} \in \mathbf{A}_u} \;\;\boldsymbol{\alpha}^\top \mathbf{K}_0 \boldsymbol{\alpha} + \boldsymbol{\alpha}^\top \hat{\mathbf{K}} \boldsymbol{\alpha} + \lambda_p\|\boldsymbol{\alpha} \|^2 - 2\boldsymbol{\alpha}^\top\mathbf{k}_0 - 2\boldsymbol{\alpha}^\top\hat{\mathbf{q}}_u
\end{align}
where both $\boldsymbol{\alpha}^\top \mathbf{K}_0 \boldsymbol{\alpha}$ and $-2\boldsymbol{\alpha}^\top\mathbf{k}_0$ are constant values independent from $\boldsymbol{\alpha}$:
$$
	\boldsymbol{\alpha}^\top \mathbf{K}_0 \boldsymbol{\alpha} = k_0\sum\limits_{i\in \mathcal{I}_u} \sum\limits_{j\in \mathcal{I}_u} \alpha_i \alpha_j = k_0\sum\limits_{i\in \mathcal{I}_u} \alpha_i \sum\limits_{j\in \mathcal{I}_u}  \alpha_j = k_0;
$$
$$
	-2\boldsymbol{\alpha}^\top\mathbf{k}_0 = -2\sum\limits_{i \in \mathcal{I}_u} k_0 \alpha_i = -2k_0 \sum\limits_{i\in \mathcal{I}_u} \alpha_i = -2k_0;
$$
and hence the solution of the optimization problem does not depend on $\mathbf{K}_0$:
$$
	\boldsymbol{\alpha}^* = \argmin\limits_{\boldsymbol{\alpha} \in \mathbf{A}_u} \;\;\boldsymbol{\alpha}^\top \hat{\mathbf{K}} \boldsymbol{\alpha} + \lambda_p\|\boldsymbol{\alpha} \|^2 - 2\boldsymbol{\alpha}^\top\hat{\mathbf{q}}_u.
$$
and it is the same as in (\ref{opt3}).
\end{proof}

For this reason we can ``sparsify'' these kernels (we will call theme RDP Kernels: Reduced Dot Product Kernels) by removing the zero degree factor obtaining kernel matrices whose sparsities depend only on the distribution of the input data since they are defined as linear combination of powers of dot products. This also implies that the sparsity of a RDP kernel is exactly the same as in the simple linear kernel  (i.e., $\mathbf{K} = \mathbf{X}^\top \mathbf{X}$).

\subsection{Sparsity and long tail distribution}
As mentioned in Section \ref{eff_cf-omd}, CF datasets are, in most of the cases, very sparse and in general the distribution of the ratings has a long tail form from the items perspective \cite{Anderson:2006}. 
This means that a small set of items, the most popular ones, receive great part of the whole set of ratings.

What we need to understand are the conditions under which a RDP kernel remains sufficiently sparse. 
To study this phenomenon we use the linear kernel, but results also apply for every other RDP kernel as well because they contain exactly the same zero entries as the linear one (which is in fact a special case of RDP kernel).\\

Let $\mathbf{K} = \mathbf{X}^\top \mathbf{X}$ ($\mathbf{X} \in \mathbb{R}^{n\times m}$) be a kernel matrix and let $\mathbb{P}(K_{ij} \neq 0)$ be the probability that the entry $K_{ij}$ is not zero. 
Given an a-priori probability distribution over the ratings, and assuming the independence of the ratings, we can estimate the probability of having a value different from zero in the kernel matrix with: 
\begin{align}\label{eq:prob}
	\mathbb{P}(K_{ij} \neq 0) &= 1 - \mathbb{P}(K_{ij} = 0) \\
	&= 1 - \prod\limits_h \mathbb{P}(x_{ih}\cdot x_{jh} = 0) \\
	&= 1 - \prod\limits_h (1 - \mathbb{P}(x_{ih}\cdot x_{jh} \neq 0)) \\
	&= 1 - \prod\limits_h (1 - \mathbb{P}(x_{ih} \neq 0)\mathbb{P}(x_{jh} \neq 0)) \\
	&= 1 - (1 - \mathbb{P}(x_{ih}\neq 0) \cdot \mathbb{P}(x_{jh}\neq 0))^n
\end{align}
where $\mathbb{P}(x_{ih}\neq 0)$ and $\mathbb{P}(x_{jh}\neq 0)$ are the probability of having a non zero entry in the rating matrix. It is worth to notice that this probability, in the uniform case, is actually the density of the matrix. However, it does not take into account the fact that all the elements in the diagonal of the kernel are for sure non zero. 
With this consideration in mind we can define an estimate of the kernel density $d(\mathbf{K})$, with $\mathbf{K} \in \mathbb{R}^{m\times m}$, as follows:
$$
	d(\mathbf{K}) = \frac{1}{m^2} \left[m + (m^2-m)\mathbb{P}(K_{ij} \neq 0) \right].
$$
 
Intuitively, we can argue that anytime both $\mathbf{x}_i$ and $\mathbf{x}_j$ are popular items, i.e., $\mathbb{P}(x_{ih}\neq 0)$ and $\mathbb{P}(x_{jh}\neq 0)$ are close to 1, then $\mathbb{P}(K_{ij} \neq 0)$ tends to be high and hence $\mathbf{K}$ is likely to be dense. On the contrary, when one of the two vectors represents an unpopular item, then the probability $\mathbb{P}(K_{ij} \neq 0)$ is likely close to zero. 
Since we are assuming a long tail distribution over the items, most of the kernel entries result from a dot product of two unpopular items with very few users that rated them and so the kernel tends to be sparse.

However, up to now, we are assuming a uniform distribution over the users and in real datasets this is often not the case. 
Any other probability distribution would highly affect our estimation $d(\mathbf{K})$. This is because, if we assume a long tail distribution over the users, the probability of having at least one user in common between two items would be generally high, since the ratings for an item are likely to be concentrated to the (few) most active users.
Considering that a mathematical proof of this intuition is quite complicated, in the next section we provide an empirical analysis on real CF datasets.


\subsubsection{Empirical analysis of CF datasets}
In order to validate our thesis, we empirically analyze a set of famous CF datasets comparing the theoretical sparsity with uniform ratings distribution with the sparsity of the linear kernel. We expect that the long tail distribution of the ratings will tend to lower the likelihood of having a non zero value in the kernel matrix, with respect to the $d(\mathbf{K})$ estimate. 

The empirical analysis has been performed as follows: for each dataset, we build the corresponding rating matrix $\mathbf{R}$, we calculate the expected density using (\ref{eq:prob})  by fixing $\mathbb{P}(x_{ih} \neq 0)$ equals to the density of $\mathbf{R}$; we calculate the linear kernel $\mathbf{K}=\mathbf{R}^\top \mathbf{R}$ and finally we compare the sparsity of the kernel with $d(\mathbf{K})$.
Table \ref{tab:sparsity} summarizes the results.
\begin{table}[htbp]
	\centering
	\begin{tabular}{|c|c|c|c|}
		\hline
		\textbf{Dataset}  & \textbf{Density} $\mathbf{R}$ & \textbf{Density} $\mathbf{K}$ & $d(\mathbf{K})$ \\ \hline
		Delicious &  0.08\% & 0.128\% & 0.13\% \\ \hline
		LastFM & 0.28\% & 0.85\% & 1.48\% \\ \hline
		Book-Crossing & 0.003\% & 0.687\% & 0.011\% \\ \hline
		Movielens 10M &  1.34\% & 85.56\% & 99.99\% \\ \hline
		Netflix & 0.99\% & 98.03\% & 99.99\% \\ \hline
		Ciao &  0.025\% & 2.51\% & 0.12\% \\ \hline
		FilmTrust & 1.13\% & 11.11\% & 17.74\% \\ \hline
	\end{tabular}
	\caption{Analysis of the sparsity of the linear kernel.\label{tab:sparsity}}
\end{table}

Although our intuition seems to work with most of the datasets, we can notice that with the \textit{Ciao} and \textit{Book Crossing} datasets the kernels are more dense than the estimation $d(\mathbf{K})$. 

\begin{figure}[h!]
\subfloat[Item Popularity]{\includegraphics[width = 2.2in]{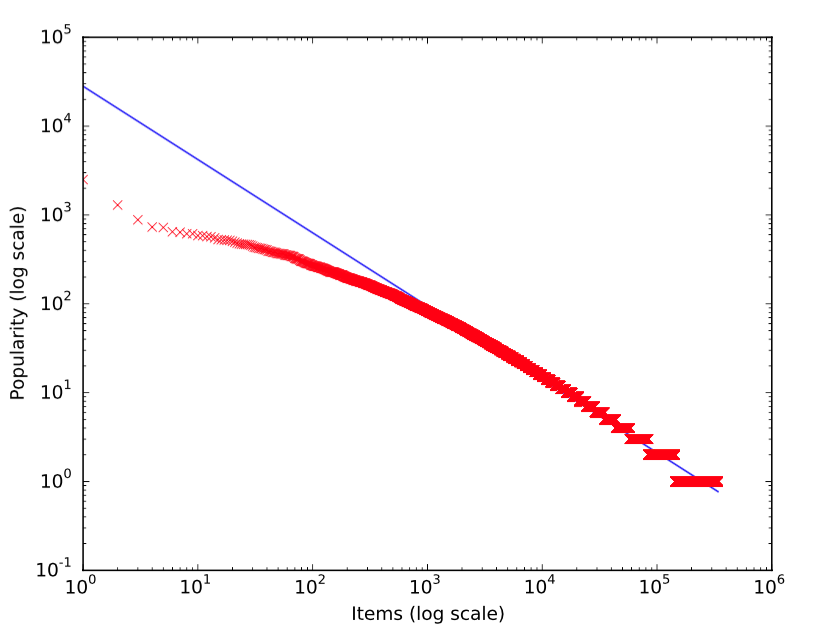}}
\subfloat[User Activity]{\includegraphics[width = 2.2in]{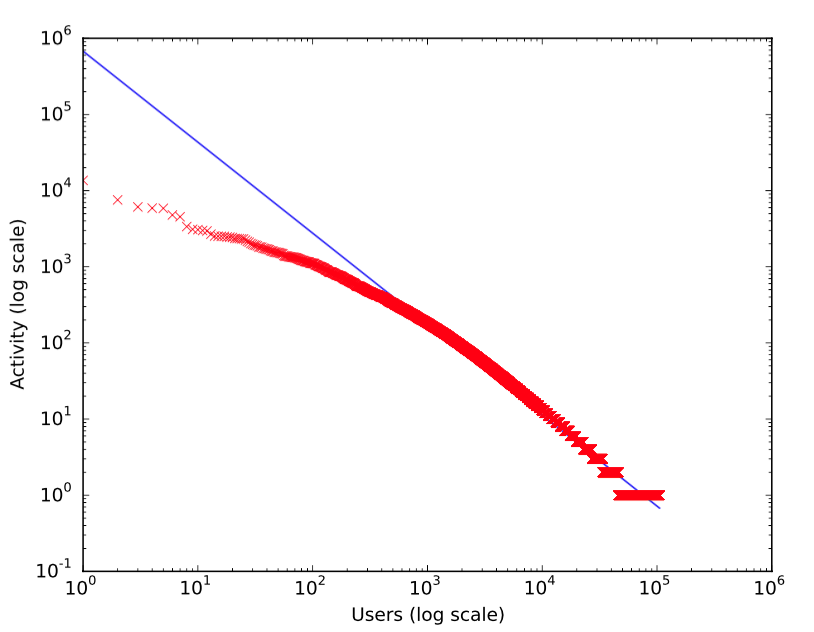}}
\caption{Book Crossing. The plots are in loglog scale.}
\label{fig:law-bx}
\end{figure}

\begin{figure}[h!]
\subfloat[Item Popularity]{\includegraphics[width = 2.2in]{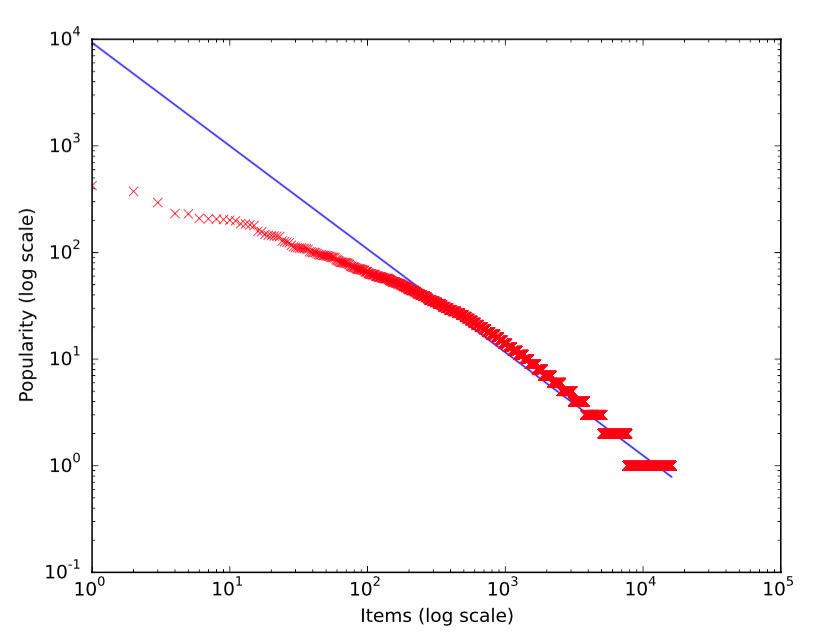}}
\subfloat[User Activity]{\includegraphics[width = 2.2in]{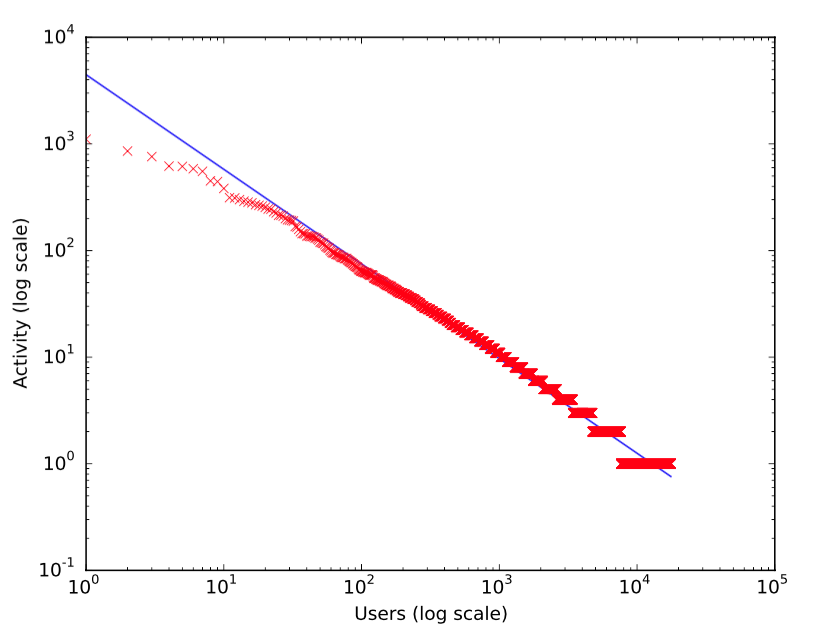}}
\caption{Ciao. The plots are in loglog scale.}
\label{fig:law-ciao}
\end{figure}

\begin{figure}[h!]
\subfloat[Item Popularity]{\includegraphics[width = 2.2in]{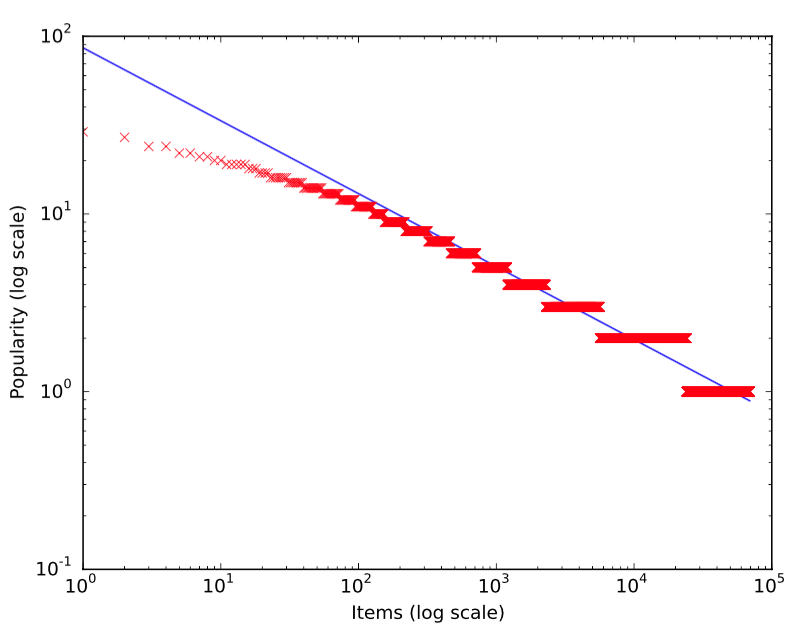}}
\subfloat[User Activity]{\includegraphics[width = 2.2in]{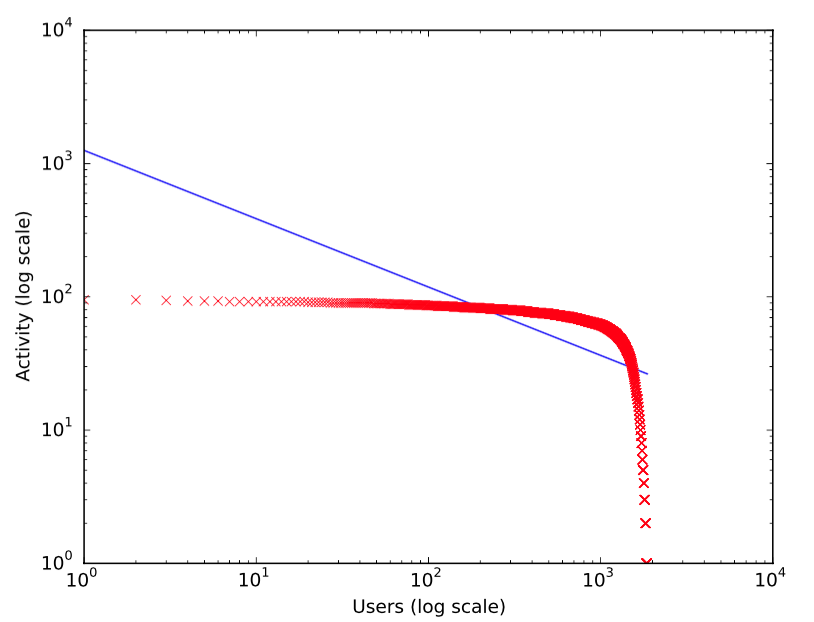}}
\caption{Delicious. The plots are in loglog scale.}
\label{fig:law-del}
\end{figure}

\begin{figure}[h!]
\subfloat[Item Popularity]{\includegraphics[width = 2.2in]{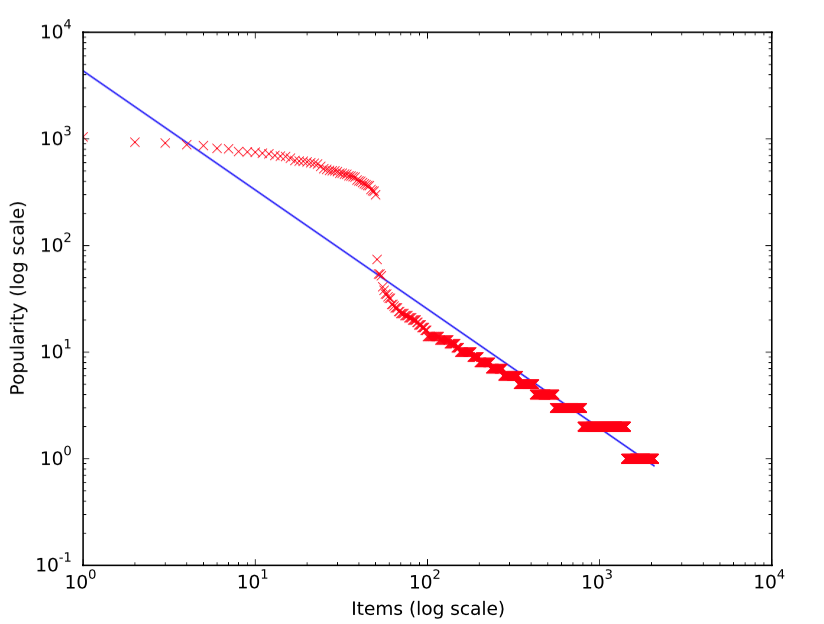}}
\subfloat[User Activity]{\includegraphics[width = 2.2in]{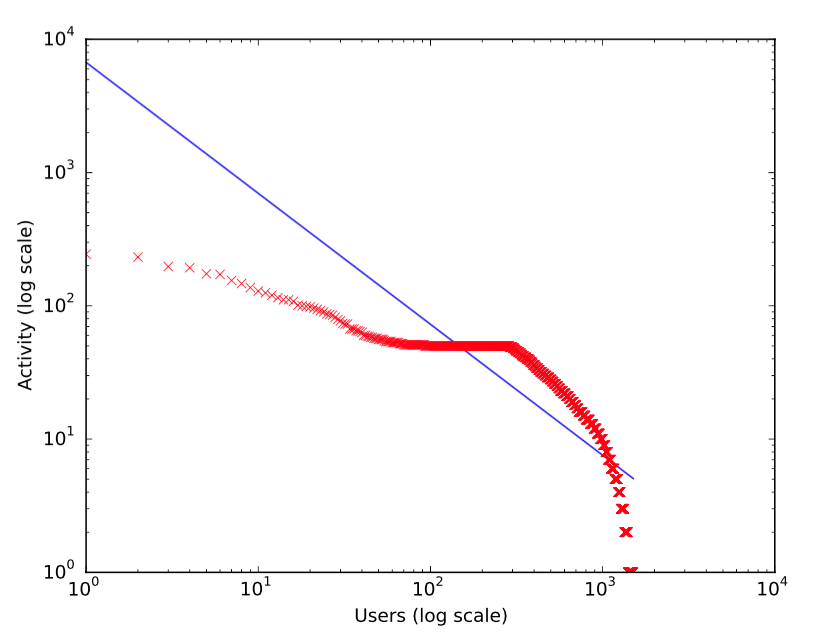}}
\caption{Film Trust. The plots are in loglog scale.}
\label{fig:law-filmtrust}
\end{figure}

\begin{figure}[h!]
\subfloat[Item Popularity]{\includegraphics[width = 2.2in]{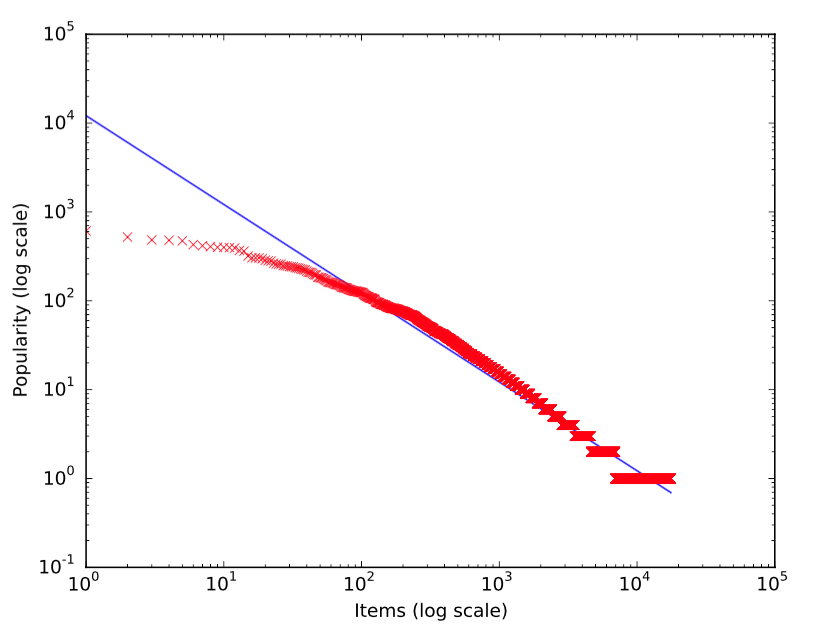}}
\subfloat[User Activity]{\includegraphics[width = 2.2in]{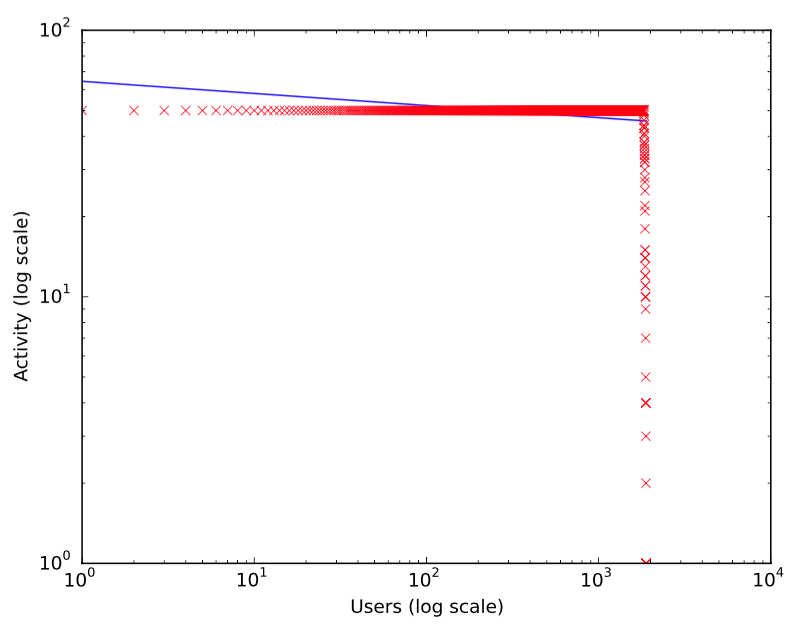}}
\caption{LastFM. The plots are in loglog scale.}
\label{fig:law-lastfm}
\end{figure}

\begin{figure}[h!]
\subfloat[Item Popularity]{\includegraphics[width = 2.2in]{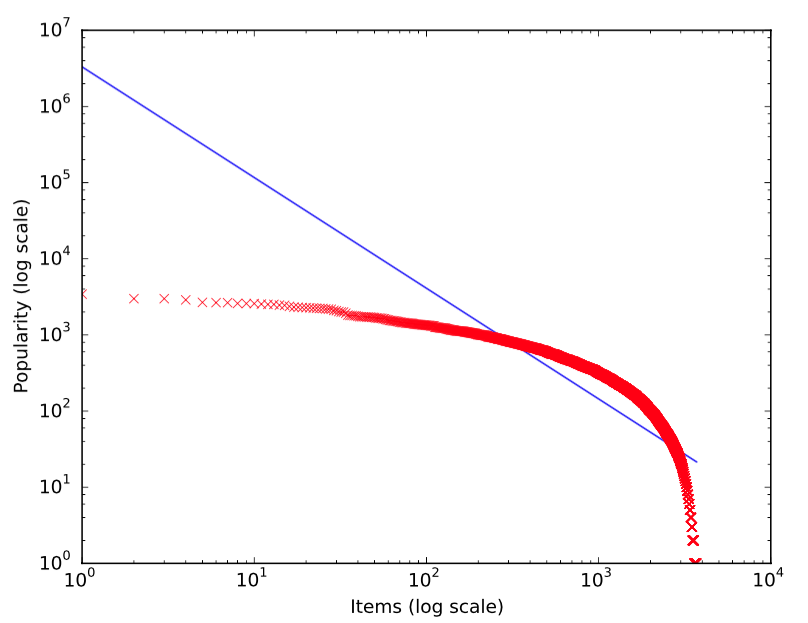}}
\subfloat[User Activity]{\includegraphics[width = 2.2in]{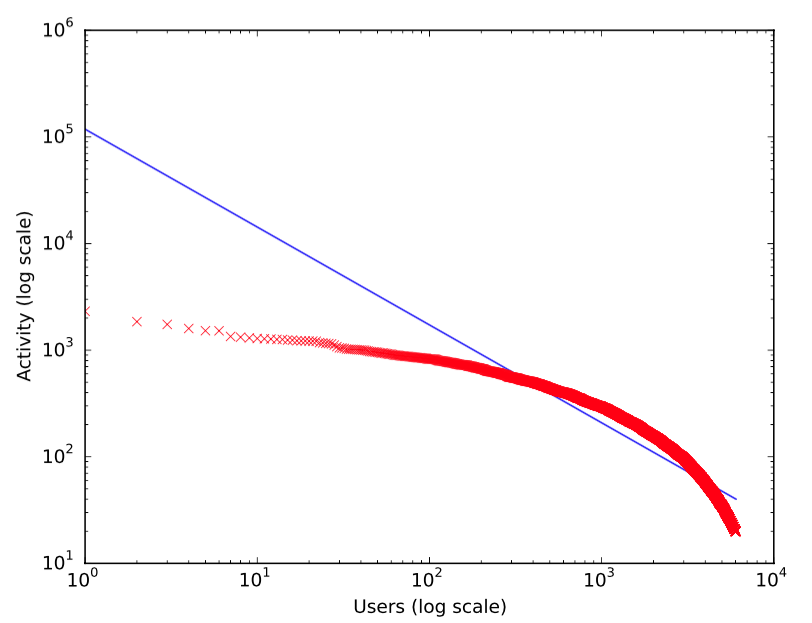}}
\caption{Movielens. The plots are in loglog scale.}
\label{fig:law-ml}
\end{figure}

\begin{figure}[h!]
\subfloat[Item Popularity]{\includegraphics[width = 2.2in]{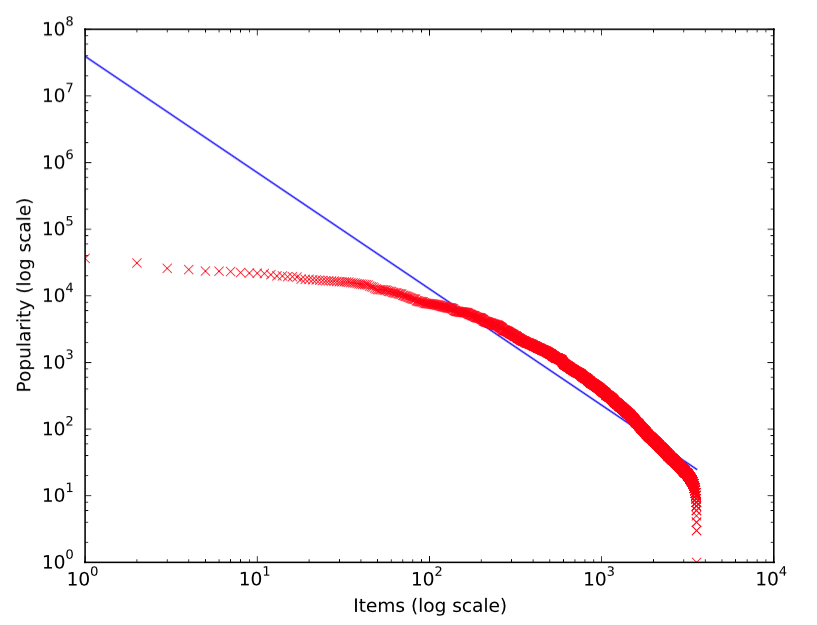}}
\subfloat[User Activity]{\includegraphics[width = 2.2in]{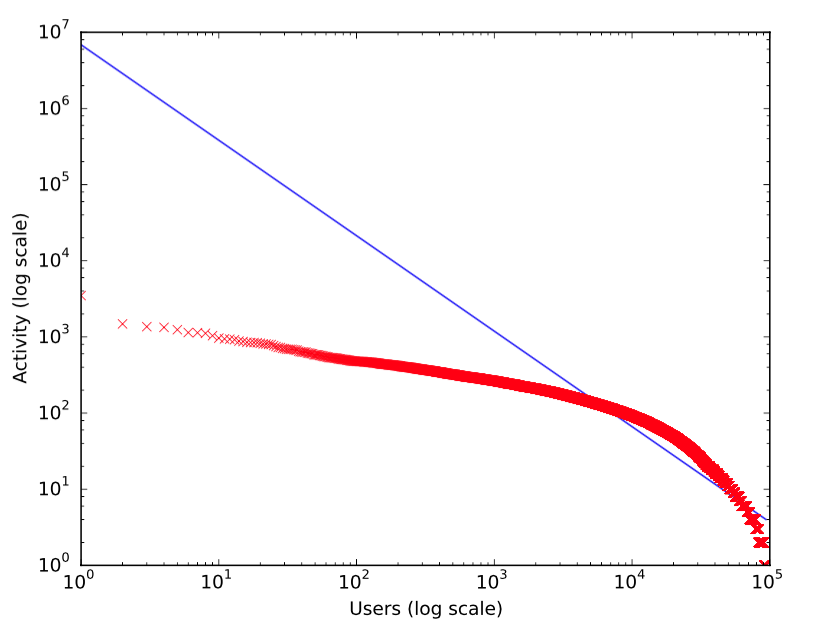}}
\caption{Netflix. The plots are in loglog scale.}
\label{fig:law-nfx}
\end{figure}

The reason why these datasets behave differently is clearly depicted in the plots \ref{fig:law-bx} - \ref{fig:law-nfx}.
Every pair of plots show the distribution of the item popularity and the user activity. The plots have a loglog scale and the blue line represents the best fitting power low function. The fitting has been made using the least square method.


From the plots we can observe that:
\begin{itemize}
	\item none of the item distributions follows exactly a power low, especially in the head of the distribution;
	\item datasets with a very dense kernel tend to have shorter tails: the right part of the plot exceed the power low line;
	\item generally items are long tailed while users tend to be more uniform;
	\item users distributions are in general not well fitted by a power law, with the exception of \textit{Ciao} and \textit{Book Crossing} datasets.
\end{itemize}

The observations listed above point out that \textit{Ciao} and \textit{Book Crossing} are the only two datasets with a well defined long tail distribution over both users and items. This confirm our intuition about the likelihood of having a denser kernel with both long tailed distributions.

In conclusion, the long tail distribution over the items keeps the RDP kernels sparse while a long tail distribution over the users increase the density. 

\subsection{Approximation of $\mathbf{q}_u$}
Using the RDP kernels, we can further optimize the complexity by providing a good approximation of $\mathbf{q}_u$ that can be computed only once, instead of $n_{\textit{ts}}$ times. The idea consists in replacing every $q_{ui}$ with an estimate of $\mathbb{E}[K(\mathbf{x}_i, \mathbf{x})]$ which is the expected value of the kernel between the item $i$ and every other items. Formally, consider, without any loss of generality, a normalized kernel function $K$ and let the approximation of $\mathbf{q}_u$ be $\tilde{\mathbf{q}}$ such that:
$$
\tilde{q}_i = \frac{1}{m} \sum_{j \in \mathcal{I}} K(\mathbf{x}_i, \mathbf{x}_j).
$$
At each component of $\hat{\mathbf{q}}$, the approximation error is bounded by $\frac{2m_u^+}{m}$, which is linear on the sparsity of the dataset.
\begin{proof}
\begin{align}
	|\hat{q}_i - q_{ui}| &= \left| \frac{1}{m} \sum\limits_{j \in \mathcal{I}} K(\mathbf{x}_i, \mathbf{x}_j) - \frac{1}{m_u^-} \sum\limits_{j \notin \mathcal{I}_u} K(\mathbf{x}_i, \mathbf{x}_j) \right| \\
	&= \left| \frac{1}{m} \left[ \sum\limits_{j \in \mathcal{I}_u} K(\mathbf{x}_i, \mathbf{x}_j) + \sum\limits_{j \notin \mathcal{I}_u} K(\mathbf{x}_i, \mathbf{x}_j) \right] - \frac{1}{m_u^-} \sum\limits_{j \notin \mathcal{I}_u} K(\mathbf{x}_i, \mathbf{x}_j) \right| \\
	&= \left| \frac{1}{m} \sum\limits_{j \in \mathcal{I}_u} K(\mathbf{x}_i, \mathbf{x}_j) - \frac{m - m_u^-}{m \cdot m_u^-} \sum\limits_{j \notin \mathcal{I}_u} K(\mathbf{x}_i, \mathbf{x}_j) \right|  \\
	&\leq \left| \frac{1}{m} \sum\limits_{j \in \mathcal{I}_u} K(\mathbf{x}_i, \mathbf{x}_j) \right| + \left| \frac{m - m_u^-}{m \cdot m_u^-} \sum\limits_{j \notin \mathcal{I}_u} K(\mathbf{x}_i, \mathbf{x}_j) \right| \\
	&\leq \left| \frac{m_u^+}{m} \right| + \left| \frac{m-m_u^-}{m \cdot m_u^-}m_u^- \right| \leq \frac{m_u^+ + m - m_u^-}{m} = 2\frac{m_u^+}{m}.
\end{align}
\end{proof}

\section{Experiments and Results}\label{result}
Experiments have been performed comparing the proposed methods against the state-of-the-art method on MSD (MSDW) with respect to the ranking quality and computational performance.
 
We also compared our framework, in terms of AUC, against other state-of-the-art methods on top-N recommendation with implicit feedback, namely WRMF and BPR.

Our framework and MSDW are both implemented in Python\footnote{We used CVXOPT package to solve the optimization problem}\footnote{The MSDW implementation is available at http://www.math.unipd.it/~aiolli/CODE/MSD/}\footnote{The framework implementation is available at https://github.com/makgyver/pyros}, while for WRMF and BPR we used the java implementation provided by the open source LibRec library \footnote{http://www.librec.net/}.

\subsection{Datasets}
In this section we introduce the datasets used in the experiments. Table \ref{tab:datasets} shows 
a brief description of the datasets.

\begin{table}[h!]
	\centering
	\begin{tabular}{|c|c|c|c|c|}
		\hline
		\textbf{Dataset} & \textbf{Item type} & $|\mathcal{U}|$ & $|\mathcal{I}|$ & $|\mathcal{R}|$\\ \hline
		MovieLens & Movies & 6040 & 3706 & 1M\\ \hline
		MSD & Music & 1.2M & 380K & 48M\\ \hline
		Ciao & General & 17615 & 16121 & 72664\\ \hline
		Netflix & Movie & 93705 & 3561 & 3.3M\\ \hline
		FilmTrust & Movie & 1508 & 2071 & 35496\\ \hline
	\end{tabular}
	\caption{Brief description of the used datasets.\label{tab:datasets}}
\end{table}

The MSD dataset is used only to demonstrate the applicability of our kernel method to huge datasets. All the other datasets are used to compare our framework with the state-of-the-art method in top-N recommendation with implicit feedback.

\subsection{Experimental setting}
Experiments have been performed 5 times for each dataset. Datasets have been pre-processed as described in the following:
\begin{enumerate}
	\item we split randomly the users in 5 sets of the same dimension;
	\item for each user in a set we further split its ratings in two halves;
	\item at each round test, we use all the ratings in 4 sets of users plus the first half of ratings of the remaining set as training set, and the rest as test set.
\end{enumerate}
This setting avoids situations of cold start for users, because in the training phase we have at least a rating for every user. We also force users with less than 5 ratings to be in the training set.
The results reported below are the averages (with its standard deviations) over the 5 folds.

\subsection{Evaluation measures}
The rankings' evaluation metric used to compare the performances of the methods is the AUC (Area Under the receiver operating characteristic Curve) defined as in the following:
$$
	\textit{AUC} = \frac{1}{|\mathcal{U}|} \sum\limits_{u\in\mathcal{U}} \frac{1}{|\mathcal{I}_u| \cdot |\mathcal{I}\setminus \mathcal{I}_u|} \sum\limits_{i \in \mathcal{I}_u} \sum\limits_{j \notin \mathcal{I}_u} \mathbb{I}[\hat{r}_{ui} > \hat{r}_{uj}]
$$
where $\mathbb{I} : \textit{Bool} \rightarrow \{0,1\}$ is the indicator function which returns 1 if the predicates is true 0 otherwise.
In the experiments which use the MSD dataset we also compared the algorithms using the same metric as in the challenge, that is the Mean Average Precision at 500 (mAP@500), which is the mean over all users of the average precision at $N$ ($N=500$). Formally is defined as:
$$
	AP(\pi_u)@N = \frac{1}{\min(|\mathcal{I}_u|, N)} \sum\limits_{k=1}^{N} P(\pi_u)@k \cdot \hat{r}_{u\pi_u(k)},
$$
where $\mathcal{I}_u$ is the set of positive associated items with the user $u$, $\pi_u$ is the items ranking for user $u$, such that $\pi_u(k) = i$ means that item $i$ is ranked at position $k$, and $P(\pi_u)@k$ is the precision at $k$:
$$
	P(\pi_u)@k = \frac{1}{k} \sum\limits_{p=1}^k \hat{r}_{u\pi_u(p)}.
$$

\subsection{Results}
\subsubsection{MSD}
We used MSD as described in the Kaggle challenge\footnote{https://www.kaggle.com/c/msdchallenge}: the training set is composed by 1M users (plus 10K users as validation set) with all their listening history and for the rest (i.e., 100K users) only the first half of the history is provided, while the other half constitutes the test set. In these experiments we fixed the $\lambda_p$ parameter to 0.01.

Results are presented in Table \ref{msd-res}. In this case MSDW maintains its record performance in terms of mAP@500, while for the AUC all methods have very good results. This underline the fact that both ECF-OMD and CF-KOMD (polynomial kernel with $c=1$) try to optimize the AUC rather than the mAP.
\begin{table}[h!]
\centering
\begin{small}
	\begin{tabular}{r|c|c|c|} 
		 \cline{2-4}
		&  MSDW $(\alpha, q)$ & ECF-OMD $(\lambda_p)$ & CF-K $(\lambda_p)$ \\ \cline{2-4}
		& $0.15, 3$ & $0.1$ & $0.1$ \\ \hline
		\multicolumn{1}{|r|}{mAP@500} & \textbf{0.16881} & 0.16391 & 0.15967 \\ \hline
		\multicolumn{1}{|r|}{AUC} & \textbf{0.97342} & 0.97034 & 0.97065 \\ \hline
	\end{tabular}
	\caption{Ranking accuracy on MSD using AUC and mAP@500.\label{msd-res}}
\end{small}
\end{table}

\begin{figure}[h!]
	\centering
	\includegraphics[scale=1]{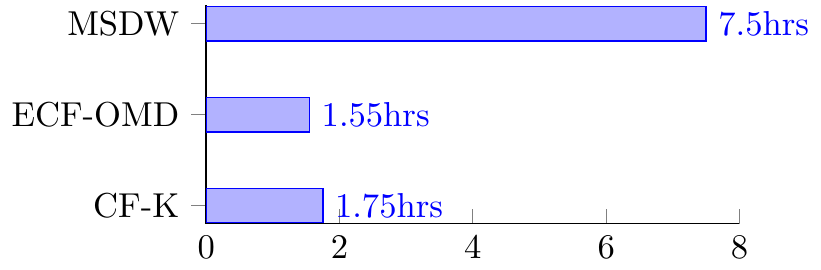}
	\caption{Average computational time in hours for 1K users.}
	\label{comp-time}
\end{figure}

\noindent The computational costs on this dataset are reported in Figure \ref{comp-time}.

The results are the average computing time over 1K test users. All methods run on a machine with 150Gb of RAM and 2 x Eight-Core Intel(R) Xeon(R) CPU E5-2680 0 @ 2.70GHz. Actually the times in Figure \ref{comp-time} have a constant overhead due to read operations. Results show that ECF-OMD and CF-K (abbreviation for CF-KOMD) are almost 5 time faster than MSDW even though they require more RAM to store the kernel matrix.
It is worth to notice that CF-K has a computational time very close to ECF-OMD, and this highlights the positive effects of the complexity optimization presented in this paper.

\subsubsection{Other datasets}

This section shows the performance for the top-N recommendation task, in terms of AUC, achieved by our framework. We compared our methods with some state-of-the-art techniques.
In particular, we used the following settings for each of the competing algorithm:
\begin{description}
	\item[ECF-OMD]: we fixed the regularization parameter $\lambda_p = 0.01$;
	\item[CF-K$_P$]: it is the CF-KOMD method with the polynomial kernel. We tested different values for $c \in \{0.5, 1, 2, 4\}$ and we fixed $\lambda_p = 0.01$;
	\item[CF-K$_T$]: it is the CF-KOMD method with the tanimoto kernel. We fixed the regularization parameter $\lambda_p = 0.01$;
	\item[MSDW]: tests have been performed varying the value of the $\alpha$ parameter in the real range [0,1] with a step of 0.25;
	\item[WRMF]: we reported only the performance achieved with $\alpha=1$ since the results obtained with different $\alpha$ were substantially the same. We fixed the maximum number of iteration to 30, the learning rate $\rho=0.001$, the regularization term $\lambda=0.001$ and the number of factor $k=100$;
	\item[BPR]: we do not have free parameter. We fixed, as in WRMF, the maximum number of iteration to 30, the learning rate $\rho=0.001$ and the regularization term $\lambda=0.001$.
\end{description}

Table \ref{tab:results} summarizes the obtained results.

\begin{table}[h!]
	\hspace*{-1.7em}
	\centering
	\begin{tabular}{r|c|c|c|c|c|}
		\multicolumn{2}{c|}{}&\textbf{MovieLens} & \textbf{Netflix} & \textbf{FilmTrust} & \textbf{Ciao} \\ \hline
		MSDW & $\alpha = 0.00$ & $0.867_{\pm0.001}$ & $0.939_{\pm0.0002}$ & $0.961_{\pm0.004}$ & $\mathbf{0.824_{\pm0.008}}$ \\
		MSDW & $\alpha = 0.25$ & $0.870_{\pm0.001}$ & $0.939_{\pm0.0002}$ & $0.960_{\pm0.005}$ & $0.813_{\pm0.01}$ \\
		MSDW & $\alpha = 0.50$ & $0.875_{\pm0.001}$ & $0.936_{\pm0.0006}$ & $0.959_{\pm0.005}$ & $0.805_{\pm0.01}$ \\
		MSDW & $\alpha = 0.75$ & $0.862_{\pm0.001}$ & $0.912_{\pm0.0001}$ & $0.955_{\pm0.005}$ & $0.793_{\pm0.008}$ \\
		MSDW & $\alpha = 1.00$ & $0.458_{\pm0.007}$ & $0.409_{\pm0.019}$ & $0.833_{\pm0.011}$ & $0.784_{\pm0.009}$ \\ \hline 
		ECF-OMD & - & $0.895_{\pm0.0003}$ & $\mathbf{0.943_{\pm0.001}}$ & $0.961_{\pm0.005}$ & $0.718_{\pm0.006}$ \\ 
		CF-$K_P$ & $c=0.50$ & $0.893_{\pm0.0003}$ & $0.937_{\pm0.001}$ & $0.961_{\pm0.006}$ & $0.731_{\pm0.003}$ \\
		CF-$K_P$ & $c=1.00$ & $0.893_{\pm0.0003}$ & $0.937_{\pm0.001}$& $0.960_{\pm0.006}$ & $0.730_{\pm0.003}$ \\
		CF-$K_P$ & $c=2.00$ & $0.894_{\pm0.0003}$ & $0.938_{\pm0.001}$ & $0.959_{\pm0.006}$ & $0.721_{\pm0.003}$ \\
		CF-$K_P$ & $c=4.00$ & $\mathbf{0.896_{\pm0.0003}}$ & $0.938_{\pm0.001}$ & $0.958_{\pm0.006}$ & $0.718_{\pm0.003}$ \\
		CF-$K_T$ & - & $0.895_{\pm0.0004}$ & $0.941_{\pm0.001}$& $\mathbf{0.964_{\pm0.005}}$ & $0.734_{\pm0.004}$ \\ \hline
		WRMF & $\alpha=1.00$ & $0.870_{\pm0.006}$ & $0.804_{\pm0.002}$ & $0.947_{\pm0.007}$ & $0.565_{\pm0.004}$ \\ 
		BPR & - & $0.854_{\pm0.005}$ & $0.843_{\pm0.001}$ & $0.954_{\pm0.008}$ & $0.549_{\pm0.004}$ \\ 
		\hline
	\end{tabular}
	\caption{AUC results of our framework against state-of-the-art methods.\label{tab:results}}
\end{table}

On the \textit{MovieLens} dataset, methods of our framework have significant better performance against all the other, achieving an AUC of 0.896 with CF-KOMD with the polynomial kernel ($c=4$). CF-KOMD has the higher AUC (0.964) also in the \textit{FilmTrust} dataset, but this time with the Tanimoto kernel. In this dataset, anyway, the polynomial achieved result comparable with ECF-OMD. Good results are also achieved by the MSDW.
On the \textit{Netflix} dataset, the best AUC is 0.943 by ECF-OMD. We got similar result with the Tanimoto kernel.
Surprisingly, with the \textit{Ciao} dataset, MSDW obtained very good result while all the other approaches are quite behind,n particular, WRMF and BPR have a very poor performances.\\

\begin{figure}[h!]
	\centering
	\includegraphics[scale=.6]{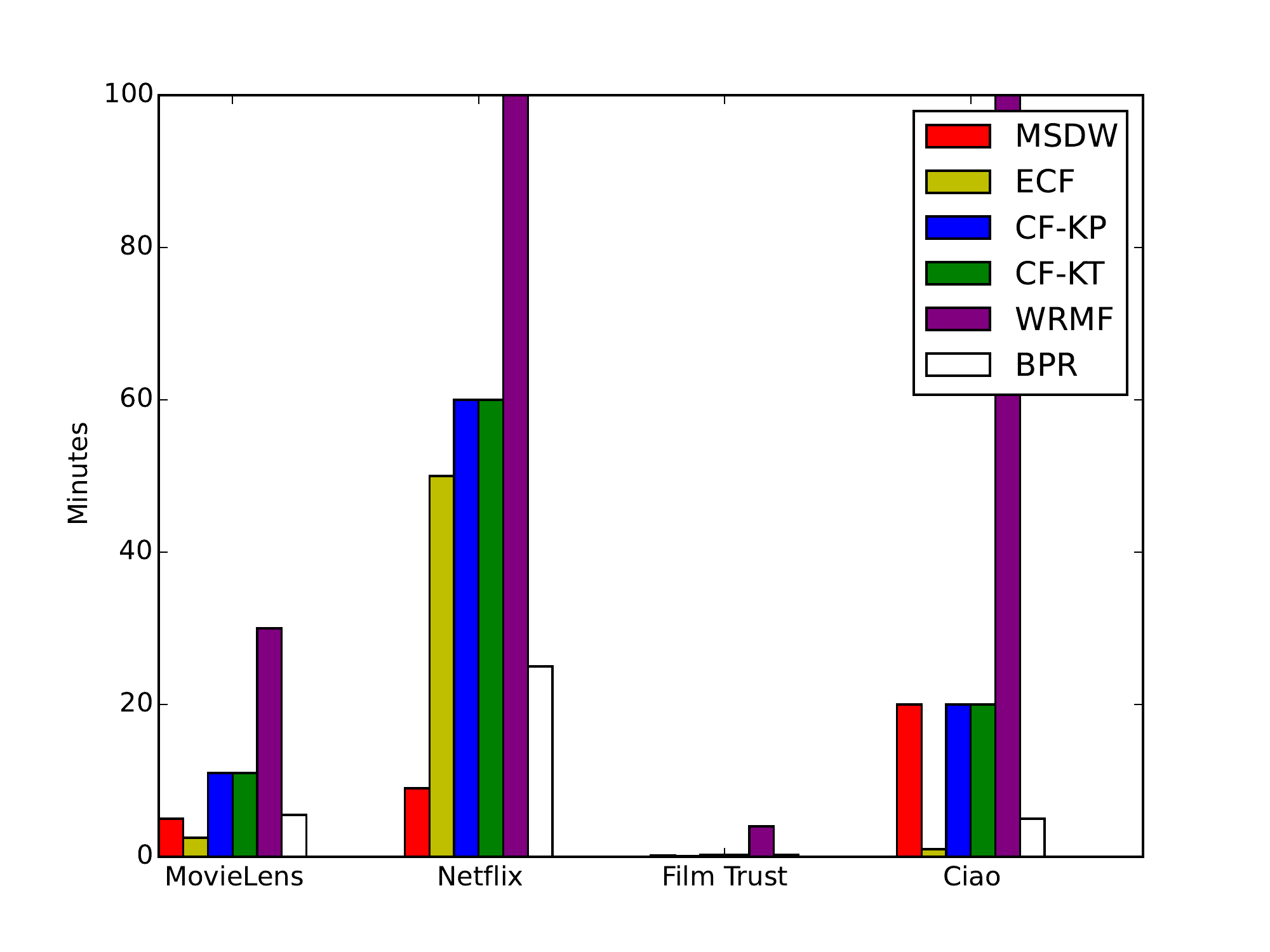}
	\caption{Execution time took by the tested methods.\label{fig:time}}
\end{figure}

We also compared the execution time of the algorithm. All these experiments has been made on a MacBook Pro late 2012 with 16GB of RAM and CPU Intel(R) Core i7 @ 2.70GHz.
The results are shown in Figure \ref{fig:time}.
The first thing we can notice is that the WRMF algorithm is always the most time consuming one. To keep the plot readable, we cut the bars longer than 100 minutes. Actually, in both \textit{Netflix} and \textit{Ciao} datasets WRMF took more than 7 hours (i.e., 420 minutes).
The two kernel based methods, CF-K$_P$ and CF-K$_T$ took almost the same time and they are a little bit slower than the linear (ECF-OMD) method which is often the fastest one. MSDW is in general quite fast, but it seems to suffer when the number of items increase (e.g., \textit{Ciao} dataset).
These results show how our framework have, both in efficacy and efficiency, performances at the state-of-the-art.

\section{Conclusions}\label{conclusion}
In this paper we have proposed a collaborative filtering kernel-based method for the top-N recommendation. The method belongs to a more general framework, inspired by preference learning and designed to explicitly maximize the AUC.
We have also proposed a strategy for the ``sparsification'' of the dot product kernels and in which conditions this strategy works.

Our analysis, conducted over CF datasets, have shown the effect of the long tail distribution on the sparsity of the kernel. 
Since this kind of distribution is very common, our results can apply in many domains other than CF.
Finally, the experiments we reported have shown that the proposed kernel-based method achieve good result in terms of AUC and it is also efficient even with large scale datasets.

\section*{Acknowledge}
This work was supported by the University of Padova under the strategic project BIOINFOGEN.





\bibliographystyle{unsrt}
\bibliography{library}



\end{document}